\documentclass[fleqn,usenatbib]{mnras}

\usepackage{newtxtext,newtxmath}
\usepackage[T1]{fontenc}

\usepackage[dvipsnames]{xcolor}
\usepackage{fontawesome}
\usepackage{graphicx}
\usepackage{xspace}
\usepackage{amsmath}
\usepackage{cancel}
\usepackage{siunitx}
\DeclareSIUnit{\yr}{yr}
\usepackage{booktabs}
\usepackage{relsize}
\usepackage{comment}
\usepackage[capitalise]{cleveref}
\usepackage[version=4]{mhchem}

\newcommand{\mtx}[2]{#1_\text{#2}}
\newcommand{\nmm}[2]{\newcommand{#1}{\ensuremath{#2}\xspace}}

\newcommand{\code}[1]{\textsc{#1}\xspace}
\newcommand{\superrad}{\code{SuperRad}}

\nmm{\GN}{G}
\nmm{\Mpl}{\mathrm{M}_\text{P}}
\nmm{\Msol}{\mathrm{M}_\odot}
\newcommand{\state}[1]{\ensuremath{\left|#1\right\rangle}\xspace}
\newcommand{\level}[3]{\state{#1#2#3}}

\newcommand{\dd}{\mathrm{d}}
\newcommand{\ee}{\mathrm{e}}
\newcommand{\ii}{\mathrm{i}}
\newcommand{\real}{R}
\newcommand{\imag}{I}
\newcommand{\nlm}{{nlm}}
\newcommand{\order}[1]{\mathcal{O}(#1)}
\newcommand{\vc}[1]{\boldsymbol{#1}}

\newcommand{\data}{\text{D}}

\nmm{\ma}{\mu}
\nmm{\fax}{f}
\nmm{\ifax}{\fax^{-1}}
\nmm{\omR}{\omega^\real}
\nmm{\omI}{\omega^\imag}
\nmm{\Mc}{\mtx{M}{cloud}}
\nmm{\Nc}{\mtx{N}{cloud}}
\nmm{\chiTop}{\mtx{\chi}{top}}
\newcommand{\abhname}{M33~X-7\xspace}
\newcommand{\smbhname}{IRAS~09149-6206\xspace}
\nmm{\astar}{a_*}
\nmm{\astarhat}{\hat{a}_*}
\nmm{\GamZero}{\Gamma_{211}}
\nmm{\GamSR}{\Gamma}
\nmm{\GamSRZero}{\GamSR_{\nlm}}
\nmm{\tauSR}{\mtx{\tau}{SR}}
\nmm{\tauEQ}{\mtx{\tau}{eq}}
\nmm{\tauBH}{\mtx{\tau}{BH}}
\nmm{\tauBose}{\mtx{\tau}{BN}}
\nmm{\tauSalpeter}{\tau_\text{Salp}}
\nmm{\tauEdd}{\tau_\text{Edd}}
\nmm{\tauAge}{\tau_\text{age}}
\nmm{\NBose}{\mtx{N}{crit}}
\nmm{\NMax}{N_{\Delta}}
\nmm{\rp}{r_+}
\nmm{\tBH}{\mtx{\tau}{BH}}
\nmm{\colfac}{\mathfrak{f}}

\newcommand{\FeKalph}{{\ce{Fe} K$\alpha$}\xspace}
\newcommand{\updated}[1]{#1}

\title[Getting More Out of ULB Constraints from BHSR]{Getting More Out of Black Hole Superradiance: a Statistically Rigorous Approach to Ultralight Boson Constraints from Black Hole Spin Measurements}

\author[Hoof et al.]{
Sebastian Hoof,\textsuperscript{1,2}\thanks{E-mail: hoof@pd.infn.it}
David J.\ E.\ Marsh,\textsuperscript{3}\thanks{E-mail: david.j.marsh@kcl.ac.uk}
J\'ulia Sisk-Reyn\'es,\textsuperscript{4,5}
James H. Matthews,\textsuperscript{6}
and Christopher Reynolds\textsuperscript{7}
\\
\textsuperscript{1}Dipartimento di Fisica e Astronomia ``Galileo Galilei,''
{Universit\`a} degli Studi di Padova, Via F.\ Marzolo 8, 35131 Padova, Italy\\
\textsuperscript{2}Istituto Nazionale di Fisica Nucleare -- Sezione di Padova, Via F.\ Marzolo 8, 35131 Padova, Italy\\
\textsuperscript{3}Theoretical Particle Physics and Cosmology, King's College London, Strand, London, WC2R 2LS, United Kingdom\\
\textsuperscript{4}Center for Astrophysics | Harvard \& Smithsonian, Cambridge, MA 02138, USA\\
\textsuperscript{5}Institute of Astronomy, University of Cambridge, Madingley Road, Cambridge CB3 OHA, United Kingdom\\
\textsuperscript{6}Department of Physics, Astrophysics, University of Oxford, Denys Wilkinson Building, Keble Road, Oxford OX1 3RH, United Kingdom\\
\textsuperscript{7}Department of Astronomy, University of Maryland, College Park, MD 20742, USA}

\date{\today}

\pubyear{2025}

\begin{document}
\label{firstpage}
\pagerange{\pageref{firstpage}--\pageref{lastpage}}
\maketitle

\begin{abstract}
Black hole (BH) superradiance can provide strong constraints on the properties of ultralight bosons (ULBs).
While most of the previous work has focused on the theoretical predictions, here we investigate the most suitable statistical framework to constrain ULB masses and self-interactions using BH spin measurements.
We argue that a Bayesian approach \updated{based on a simple timescales analysis} provides a clear statistical interpretation, deals with limitations regarding the reproducibility of existing BH analyses, incorporates the full information from BH data, and allows us to include additional nuisance parameters or to perform hierarchical modelling with BH populations in the future.
We demonstrate the feasibility of our approach using mass and spin posterior samples for the X-ray binary BH M33~X-7 and, for the first time in this context, the supermassive BH IRAS~09149-6206.
We explain the differences to existing ULB constraints in the literature and illustrate the effects of various assumptions about the superradiance process (equilibrium regime vs cloud collapse, higher occupation levels).
As a result, our procedure yields the most \updated{statistically} rigorous ULB constraints available in the literature, with important implications for the quantum chromodynamics (QCD) axion and axion-like particles.
We encourage all groups analysing BH data to publish likelihood functions or posterior samples as supplementary material to facilitate this type of analysis, \updated{and for theory developments to compress their findings to effective timescale modifications}.~\href{https://github.com/sebhoof/bhsr}{\faGithub}
\end{abstract}




\section{Introduction}

This paper will introduce a simple yet rigorous statistical framework for constraining ultralight bosons (ULBs) using black hole (BH) superradiance (SR) from observational BH spin estimates. The framework introduces no new physics, using only previously developed models of BHSR. Our approach simplifies the constraint setting procedure to use only a timescales analysis, rather than simulating BH populations and evolution. We argue that this simplified approach is appropriate given the inherent astrophysical uncertainties/systematics, and is user-friendly since the analysis is easy to repeat with modified timescale assumptions. What we develop is a rigorous Bayesian framework to set limits on ULBs that allows to simply and directly use the posterior probabilities on BH model parameters published by observers, and advocate for future publication of BH results in this format. The robust approach to data products will allow for transparent and easy compilation of future large BH datasets. Future improved understandings to the physics of BHSR should be incorporated via modifications to the characteristic timescales involved, and we demonstrate this by comparing the bosenova model to the equilibrium model for ULB self-interactions.

\citet{2010PhRvD..81l3530A,1004.3558,1411.2263} showed that BHs with large spins can probe the existence of new bosonic degrees of freedom via  BHSR.
Due to a link between the BH mass and the boson mass, the resulting constraints are relevant for ULBs.
Indeed, \citet{1411.2263,1801.01420,1805.02016,2009.07206,2011.11646,2103.06812} placed constraints on ULBs with masses $\ma \lesssim \SI{e-10}{\eV}$ using BHSR from BH spin measurements.

Ultralight bosons are of particular interest in particle astrophysics at present.
The quantum chromodynamics (QCD)~axion \citep{Peccei:1977hh,Peccei:1977ur,Wilczek:1977pj,Weinberg:1977ma}, an excellent dark matter (DM) candidate~\citep{Preskill:1982cy,Abbott:1982af,Dine:1982ah,Turner:1983he,Turner:1985si}, is among this class of particles.
There is a far-ranging experimental programme underway \citep[see e.g.][]{Marsh:2015xka,1801.08127,Chadha-Day:2021szb,Semertzidis:2021rxs,Adams:2022pbo,OHare:2024nmr} to search for QCD~axions or axion-like particles (ALPs), which can also be DM candidates \citep[see e.g.][]{1201.5902}.
As we will review later, BHSR in the stellar mass range places constraints on the lower end of the allowed \ma~range of the QCD axion.
The `fuzzy' DM model is also in the ULB class, and it is of interest as an alternative to cold DM with distinctive phenomenology on sub-galactic scales~\citep{Khlopov:1985fch,Hu:2000ke,Marsh:2013ywa,Schive:2014dra,Hui:2016ltb}, and as the DM candidate with the absolute lowest allowed mass \citep{AlvesBatista:2021eeu,Zimmermann:2024xvd}.
A combination of cosmological probes and galactic astrophysics probe the fuzzy DM mass from below, while BHSR in the supermassive mass range constrains it from above~\citep{Marsh:2015xka,Marsh:2021jmi}.
String/M-theory also predicts the existence of ULBs covering a wide mass range~\citep{Svrcek:2006yi,Conlon:2006tq,Cicoli:2012sz,2010PhRvD..81l3530A,Acharya:2010zx}, and it has recently become possible to use BHSR to rule out certain extra dimensional geometries, thus placing observational constraints on the string landscape~\citep{2103.06812}.

While BH spin estimates are a well-established probe of ULBs, there are a number of issues and subtleties related to the derivation of the constraints just discussed.
Extracting the BH mass $M$ and (dimensionless) spin \astar from BH data requires an involved analysis \citep{mclure2002,casares2014,2011.04792,2011.08948}, making this a highly specialised task where data sets need to be dealt with on a case-by-case basis.
The resulting $(M,\astar)$ estimates can be correlated and, since the strongest constraints come from BHs with close-to-maximal $|\astar|$, a standard Gaussian approximation may be inadequate.
Moreover, the theoretical computation of BHSR-related effects is challenging and depends on the BH evolution.
These complications require some compromises between a statistically rigorous analysis and the practical challenges: we wish to include uncertainties while -- at the same time -- leverage as much information as possible from the data, BH evolution, and BH population parameters.

We argue that a Bayesian framework admits all of these desiderata.
We demonstrate the feasibility of our approach based on the X-ray binary BH \abhname and the supermassive BH (SMBH) \smbhname.
These two sources were selected due to their well-understood nature and the availability of reliable $(M,\astar)$ estimates.
Ultimately, however, it is important to recognise that any BH mass and spin measurements -- regardless of the BH system or measurement technique utilised -- may be used for our method.

We use the $(M,\astar)$ distributions from the two BHs above to obtain constraints on ULB parameters.
In particular, these are the first constraints from \smbhname.
We compare the ULB limits arising from the full Bayesian treatment to other approximate statistical methods to explain the differences which arose in past works, and summarise strengths and weaknesses of each approach.
While not the focus of this work, we also discuss the role of different computations of the BHSR rate -- both with and without the inclusion of self-interactions.

At this point, let us emphasise that BH spin measurements are not the only way to search for ULBs using BHSR.
Moreover, BH spins do not, individually, offer a route to discovery, except for in the case of population statistics~\citep{1004.3558,1604.03958}.
Routes to discovery via BHSR can be pursued via dynamical means from either gravitational waves~\citep[e.g][]{1909.08854,2111.15507,KAGRA:2022osp,Zhu:2020tht}, binary orbits~\citep[e.g.][]{1804.03208,Baumann:2022pkl,Xie:2022uvp,Hannuksela:2018izj}, or direct signatures of the ULB cloud~\citep[e.g][]{GRAVITY:2023cjt}.
These methods require significantly more involved modelling, and thus more inbuilt assumptions compared to constraints from BH spin timescales, but the potential for discovery warrants the additional detailed treatment.
Any alternative searches for ULBs through BHSR are complementary to the constraints we set here.
In particular, our constraints would add additional information to the fitted ULB parameters in case of a discovery, given that the ULB's existence has to be consistent with all available data.

The goal of the present work is thus to provide rigorous and transparent BHSR constraints from observational BH spin estimates, which can be used with confidence by the community.
Explaining the differences between past works contributes to understanding their underlying assumptions and statistical methodology, and hopefully encourages more groups analysing BH data to make their posterior samples or full analysis pipelines available.
To facilitate this process, our code and the associated, redistributed BH data are publicly available~\citep{github_bhsr}.

\section{Black Hole Evolution including BHSR}\label{sec:bhsr}

Superradiance (SR) arises due to an instability of the wave equation for a massive scalar field $\phi$ with mass \ma,
\begin{equation}
    \Box \phi - \ma^2\phi = 0\, ,
    \label{eq:klein-gordon}
\end{equation}
where we initially neglect self-interactions (introduced in \cref{sec:self_interactions}).
The D'Alembertian is computed from the Kerr metric \citep{1963PhRvL..11..237K}, for which we use the `mostly positive' signature.
The SR instability arises due to the presence of growing modes for $\phi$, which can be identified as imaginary eigenvalues of the associated {Schr\"odinger} equation.
The energy required to create the `boson cloud' is extracted from the mass and spin of the BH.
The evolution of the mass and spin are treated quasi-statically, i.e.\ there is no backreaction on the Kerr spacetime and one simply evolves the parameters in the Klein-Gordon equation \eqref{eq:klein-gordon}.\footnote{Backreaction is difficult to treat even in numerical relativity for a real scalar field due to the hierarchy of timescales between the SR rate and the scalar field energy density free evolution time.
This problem is not present for vector fields, which also undergo SR but have a slower evolution time for the free field energy density \citep[see e.g.][]{1704.04791}.}

The SR rate \GamSR depends on \ma and is approximately maximised when the dimensionless `gravitational coupling' of the atom-like, bosonic states around a spinning BH,
\begin{equation}
    \alpha = \GN M \ma = \num{0.75} \left(\frac{M}{10\,\Msol}\right) \left(\frac{\ma}{\SI{e-11}{\eV}}\right) \, ,\label{eq:alpha}
\end{equation}
is $\alpha \sim 1$, i.e.\ it is approximately maximised when the boson Compton wavelength is of order the gravitational radius.
Here, \GN is Newton's constant.
While we sometimes include the Planck mass $\Mpl = 1/\sqrt{\GN} = \SI{1.22e19}{\GeV}$ as a reference scale, we will generally set $c = \GN = 1$ in what follows.

Superradiance causes the spin \astar (and mass $M$) of a BH to evolve in time,
\begin{equation}
    \dot{a}_* = - \GamSR \Mc / \omR \approx - \GamSR \Nc \, ,
\end{equation}
where the dimensionless BH spin is $\astar \equiv a/M = J/M^2$, \omR is the ULB energy (the real part of the ULB frequency), and \Mc and \Nc respectively are the total mass and occupation number of the boson cloud forming around the BH.
The latter evolves as
\begin{equation}
    \dot{N}_\text{cloud} = \GamSR \Nc + \mtx{P}{GW}/\omR \, , \label{eq:could_growth}
\end{equation}
where $\mtx{P}{GW}$ is the small power radiated in gravitational waves~(GW) from the cloud.
We neglect $\mtx{P}{GW}$ in this work and refer the reader to \citet[Sec.~6.2.2]{1501.06570} and \citet{1604.03958,Siemonsen:2022yyf} for further details on the phenomenology of GW emission from boson clouds.

In the non-relativistic regime, the boson cloud can be thought of as levels in hydrogen atom-like states.
The complete system is described by a set of coupled rate equations for the occupation numbers of each level, which grow due to BHSR, along with interactions between the levels due to scattering, and losses due to emissions to infinity and flux across the horizon \citep[see e.g.][]{2011.11646}.
An astrophysical BH will also be governed by other processes, having some typical timescale \tauBH, which we discuss further in \cref{sec:timescales}.

\subsection{Computing the free BHSR rates}\label{sec:rates}

Starting from \citet{Teukolsky:1972my}, the computation of the BHSR rates has seen a number of advances and improvements.
In particular the analytical approximations of \citet{Detweiler:1980uk} provide a straightforward way to compute BHSR rates.\footnote{After correcting a missing factor~2 \citep[e.g.][]{1209.0773,1209.4211}.}
Semi-analytical results, e.g.\ by \citet{0705.2880}, allowed a further refinement of the (semi-)analytical computations~\citep[see][]{1804.03208,1908.10370,2201.10941}.

It is illuminating to discuss the approximate analytical result.
We use the metric of an uncharged, rotating BH with angular momentum~$a$ and mass~$M$ \citep[Kerr BH;][]{1963PhRvL..11..237K}, and the coordinate system of \citet{1967JMP.....8..265B}.
The outer horizon of a Kerr BH is $\rp/M \equiv 1  + (1 - \astar^2)^{1/2}$.

The bosonic state around the BH can be labelled by `quantum numbers' \state{\nlm}, where $n \geq 2$ is the principal quantum number,\footnote{Note that some authors use $n' = n - l - 1 \geq 0$ instead.} and has the complex-valued frequency $\omega_\nlm = \omega^R_\nlm + \ii \, \omega^I_\nlm$.
The real part $\omega^R_\nlm$ corresponds to the ULB energy and the imaginary part $\omega^I_\nlm$, if positive, \updated{sets the growth rate of the field amplitude. The BHSR rate is defined as the growth of the particle number and is thus $\Gamma_\nlm=2\omega^I_\nlm$}. 
The latter is the case, and the \state{\nlm} level is superradiant, if the SR condition is fulfilled,
\begin{equation}
    \alpha / l \leq 1/2 \, . \label{eq:sr_cond}
\end{equation}
\cite{1606.02306,2405.01003} derived general bounds on \ma for superradiant states, which are also valid for levels with $n \gg 1$.

\subsubsection{Non-relativistic approximation}\label{sec:nra}

To first order in the non-relativistic approximation (NRA), we have~\citep{1004.3558}
\begin{align}
    \omega^R_\nlm &= \ma \left[1 - \frac{\alpha^2}{2n^2} + \order{\alpha^4}\right] \label{eq:nra_R} \, , \\
    \quad \omega^I_\nlm &\approx \ma \left(m\astar - 2 \rp \ma \right) \, C_{nl} \, \Pi_{lm} \, \alpha^{4l+4}\label{eq:nra_I} \\
    \text{with} \quad C_{nl} &= \frac{ 2^{4l+2} \, (n+l)! }{ n^{2l+4} \, (n-l-1)! } \left[ \frac{l!}{(2l)! (2l+1)!} \right]^2 \\
    \text{and} \quad \Pi_{lm} &= \prod_{k=1}^l \left[ k^2 \, \mathlarger(1 - \astar^2\mathlarger) + \mathlarger(m\astar - 2 \rp \ma \mathlarger)^2  \right] \, .
\end{align}
Terms at higher orders, which also introduce a dependence on the $l$ and $m$ quantum numbers, are e.g.\ discussed by \citet{1908.10370}.

\subsubsection{Higher-order corrections and continued fraction method}\label{sec:bhsr_rate_computation}

We can further refine the NRA result from \cref{sec:nra}.
For instance, the \code{Python} package \superrad~\citep{Siemonsen:2022yyf} uses the \code{qnm} package \citep{Stein:2019mop} to solve the radial Teukolsky equation using Leaver's continued fraction method \citep[CFM;][]{Leaver:1985ax} in order to compute the relativistic BHSR rates.
The numerical solution in the relativistic regime is valid at large $\alpha$ and is matched consistently onto the Newtonian approximation at small $\alpha$, found from solving for hydrogen atom-like bound states and including terms up to $\mathcal{O}(\alpha^5)$ \citep[see][]{1908.10370}.
\superrad is fast and also has good control over the numerical accuracy of the methods, with a relative error on the SR timescale estimated to be better than 1\%.
It also includes gravitational wave emission.

However, the limitation of \superrad is that it includes only modes with $l = m = 1, 2$, does not include self-interactions, and will not return rates at the highest \astar values.
We thus use our own implementation of the CFM, following \citet{0705.2880}.

\begin{figure}
    \centering
    \includegraphics[width=3.38in]{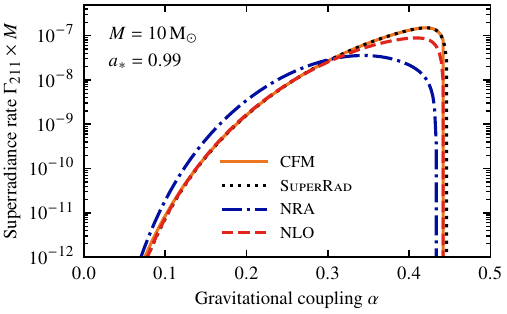}
    \caption{The \level211 superradiance rate (times the gravitational radius, $M$), computed using different methods. For benchmark values of $M = 10\,\Msol$ and $\astar = 0.99$, we compare the continued fraction method (CFM; orange line), adopted in this work, to the \superrad code (dotted, black line), and our implementation of the next-to-leading-order corrections (NLO; dashed, blue line). For completeness, we also show the non-relativistic approximation (NRA; dashed-dotted, red line).}
    \label{fig:bhsr_rates}
\end{figure}

\Cref{fig:bhsr_rates} shows the results of our implementation for the \level211 rate \GamZero (orange line), which compares well with the result of \superrad (dotted, black line).
To facilitate the numerical root-finding for the complex-valued root $\omega_{211}$, we also compute the next-to-leading-order (NLO) corrections \citep{2201.10941} as a starting point, which we show as a dashed, red line for comparison.
The computed rates differ at most by an order of magnitude, which, however, should be kept in mind when comparing different results in the literature.
This applies in particular to the NRA (dashed-dotted, blue line).

\updated{Since the present work was first drafted, a study by \cite{Witte:2024drg} appeared, which addresses the relativistic corrections in more detail. For the dominant \level211 rate, the corrections are shown to be of order 20\%. It would be possible to update the rates used in our public code to include these corrections, but at the present accuracy we believe it would have a small effect. This conclusion is validated by the agreement between the limits we set with our Bayesian simplified timescale comparison and the frequentist full trajectory analysis of  \cite{Witte:2024drg}.}~\footnote{\updated{\cite{Witte:2024drg} state that there appears to be an error in our analysis in the limit set at the low mass end. However, we have performed many consistency checks and believe the factor of two disagreement at low mass is entirely due to the statistical question being asked, i.e.\ Bayesian versus frequentist statistics.}}

\subsection{Evolution of the free field}\label{sec:free_evo}

Due to \cref{eq:could_growth}, the SR instability causes the occupation number to grow exponentially, $\Nc \propto \ee^{\GamSRZero t}$, where \GamSRZero denotes the ``free'' SR rate $\omega^I_\nlm$ from \cref{eq:nra_I}.
Eventually, however, the SR condition in \cref{eq:sr_cond} ceases to be valid and the cloud growth stops -- even in the absence of self-interactions.
By this time, the cloud will have extracted some spin $\Delta \astar$ from the BH, which can be related to the cloud occupation number $\NMax$ at the time when an amount of spin $\Delta a_*$ has been extracted, assuming negligible initial occupation number of the cloud \citep[named $\mtx{N}{max}$ in Eq.~(8)]{1411.2263},
\begin{equation}
    \NMax \approx \num{e76} \left(\frac1m\right) \left(\frac{\Delta\astar}{0.1}\right) \left(\frac{M}{10\,\Msol}\right)^2 . \label{eq:nmax}
\end{equation}
\Cref{eq:nmax} allows us to estimate whether or not a given cloud occupation level can be significantly reduce the observable spin of a BH.
In particular, if we consider the exponential growth of the cloud for the \state{nlm} level, some BH timescale \tauBH (further discussed in \cref{sec:timescales}), and the SR timescale $\tauSR = \GamSRZero^{-1}$, we can only obtain constraints on ULB properties if
\begin{itemize}
    \item \state{nlm} is superradiant according to \cref{eq:sr_cond},
    \item SR is fast enough to spin down the BH, $\tauSR < \tauBH/\ln(\NMax)$. 
\end{itemize}
Note that the typical value of $\Delta\astar$ adopted in \cref{eq:nmax} is of order the size of the observational errors on \astar.
Since the condition on \tauSR only logarithmically depends on \NMax, and thus $\Delta\astar$, even an order of magnitude difference in \NMax would not significantly change the derived limits.
Still, since the choice of $\Delta\astar$ is somewhat arbitrary, the BH evolution should be treated in more detail.
This treatment goes beyond the present analysis, where we only compare timescales.

\subsection{Self-interactions}\label{sec:self_interactions}

The effect of boson self-interactions has been considered in detail for the case of the dimensionless coupling $\lambda$ from a $\phi^4$ terms in the Lagrangian,
\begin{equation}
    \lambda \equiv \frac{\ma^2}{4!} \fax^{-2} \, ,\label{eq:self_coupling}
\end{equation}
where \fax is the ULB decay constant, which is typically connected to the energy scale of new physics leading to the ULB's existence (such as symmetry breaking or extra dimensions), while \ifax typically sets the scale of the ULB's couplings to Standard Model particles.
For a generic pseudo-Goldstone boson, \cref{eq:self_coupling} arises from Taylor expanding a generic potential with respect to the field periodicity~\fax.

For a given value of~\ma, experiments and astrophysical searches for ULBs typically place \emph{upper} limits on the couplings or \ifax, while BHSR places \emph{lower} limits on \ifax, making the methodologies highly complementary.
In the case of fuzzy DM, the relic abundance scales with~$\fax^2$ \citep{Turner:1983he,Turner:1985si,Marsh:2015xka}, with future cosmological probes reaching $\fax \sim \SI{e16}{\GeV}$ \citep[see e.g][]{Bauer:2020zsj,Farren:2021jcd,Dvorkin:2022bsc}, which overlaps with the BHSR sensitivity.

For QCD~axions, the term in \cref{eq:self_coupling} arises from a Taylor expansion of the scalar potential in chiral perturbation theory, and \citet{1511.02867} found that $\lambda = \num{-0.346(22)}$.
This prediction is possible since QCD~axion models link \ma and \fax,
\begin{equation}
    \ma = \frac{\chiTop^2}{\fax} = \SI{5.691(51)}{\nano\eV} \left(\frac{\SI{e15}{\GeV}}{\fax}\right) \, , \label{eq:qcd_axion_mass}
\end{equation}
where the QCD topological susceptibility \chiTop has been computed at NNLO \citep{1511.02867,1812.01008}.
Note that these relations may not hold for a generic ALP.
As we now discuss, self-interactions affect the SR rates.

\subsubsection{The equilibrium regime}\label{sec:equilibrium}

Initially, the ULBs will be near the vacuum and the rate \GamSRZero is the one for free bosons, as computed in \cref{sec:rates}. 
If the SR rate is faster than the inverse of the relevant BH timescale $\tauBH^{-1}$, the boson cloud can grow (see \cref{sec:free_evo}).
Its subsequent evolution is determined by the strength of ULB self-interactions, for which there are two possibilities.

The first possibility is that the cloud enters an \emph{equilibrium} between BHSR losses and level transitions caused by self-interactions and gravity~\citep{2011.11646}.
The BHSR rate in the equilibrium regime is effectively reduced compared to the vacuum rate by a factor of $\mtx{\eta}{eq} < 1$ for the equilibrium number density of bosons in the most superradiant level.
In particular, the spin-down timescale in the equilibrium regime is given by
\begin{equation}
    \tauEQ = \frac{1}{\mtx{\eta}{eq} \GamSRZero} \, , \label{eq:tau_eq}
\end{equation}
where the factor $\mtx{\eta}{eq}$ can be computed following the procedure of \citet{2011.11646}, who consider (transitions between) the \level211 and \level322 levels,
\begin{align}
    \mtx{\eta}{eq} &= \frac{2\sqrt{3}}{3} \frac{\sqrt{\GamZero \Gamma_{322\times322}^{211\times\infty}}}{\Gamma_{211\times211}^{322\times\text{BH}}} \, , \label{eq:mod_neq} \\
    \Gamma_{322\times322}^{211\times\infty} &\approx \num{1.1e-8} \alpha^8 \ma \fax^{-4} \, , \\
    \Gamma_{211\times211}^{322\times\text{BH}} &\approx \num{4.3e-7} \Big(1 + \sqrt{1 - \astar^2}\Big) \,  \alpha^{11} \ma \fax^{-4} \, ,
\end{align}
where $\alpha$ is assumed to be sufficiently small and where \fax is the ULB decay constant in \cref{eq:self_coupling}.\footnote{We will only consider the \level211 SR rate (and transitions to the \level322 level) for the equilibrium regime, making this a good assumption.}

Given values for $(M,\astar)$, we may exclude all ULB models for which, in addition to the conditions in \cref{sec:free_evo},\footnote{The logarithm of $\NMax$ in \cref{eq:nmax} is of similar size as the occupation number typically required to reach the equilibrium regime in the first place \citep{2011.11646}. We thus need to include such as condition to use the effective equilibrium rate in \cref{eq:tau_eq}.} the reduced SR rate in equilibrium is still faster than the typical BH timescale.
This can be encoded in the condition $\tauEQ < \tauBH$.

\subsubsection{Bosenovae}\label{sec:bosenovae}

The other possibility for the evolution is that a \emph{bosenova} occurs~\citep{1004.3558,Kodama:2011zc,1203.5070,1411.2263,1505.00714}.
This happens when the occupancy of the cloud reaches a critical value $\NBose$, given by
\begin{equation}
    \NBose \approx  \num{e78} \, c_0 \, \left(\frac{M}{10\,\Msol}\right)^2 \left(\frac{f}{\Mpl}\right)^2 \, \frac{n^4}{\alpha^3} \, , \label{eq:nbose}
\end{equation}
where $c_0 \approx 5$ \citep[Eq.~(9)]{1411.2263}.
Such an expression can be derived by comparing the order of magnitude size of different terms in the action \citep[e.g.][]{2103.06812}.
Thus, in a single superradiant timescale \tauSR that would otherwise lead to a maximum occupancy $\NMax$, as given by \cref{eq:nmax}, a number of bosenovae $\NBose/\NMax$ occur.
Since in each cycle the maximum occupancy is $\NBose$, the overall spin-down rate is reduced.
This can be interpreted as an increase in the relevant timescale \citep{1411.2263}:
\begin{equation}
  \tauBose = \frac{1}{\GamSRZero} \frac{\NMax}{\NBose} \ln(\NBose) \, . \label{eq:tau_bn}
\end{equation}

Given values for $(M,\astar)$, we may exclude all ULB models for which, in addition to the conditions in \cref{sec:free_evo}, the reduced SR rate from successive bosenovae is fast enough to reduce the spin.
This can be written as $\tauBose < \tauBH$.

\subsubsection{Bosenova or Equilibrium?}\label{sec:bn_vs_eq}

In both the bosenova and equilibrium scenarios, bosonic self-interactions slow down BHSR.
In the case of a bosenova, the cloud is prevented from reaching a high occupancy and exponential growth over the entire SR timescale and instead goes through many short periods where a smaller amount of spin is extracted.
Stronger self-interactions in this case decrease the value of \NBose and make each cycle shorter, thus extracting less spin overall.
In the equilibrium case, stronger self-interactions lead to more rapid boson scattering in superradiant levels, consequently reducing their equilibrium number density, and thus their ability to extract spin from the BH.

\citet{2011.11646} approximated the full superradiant evolution by including the most relevant levels and found that an equilibrium is typically reached before a bosenova occurs.
This finding was also confirmed by \citet{Omiya:2022gwu}, \updated{who included relativistic effects}.
More concretely, for $\alpha \lesssim 0.2$, \citet{2011.11646} found that a two-level description is sufficient to model the BH evolution, the inclusion of higher levels does not affect their results, and a bosenova does not occur.
For $\alpha \lesssim 0.3$, they found an equilibrium for the three-level system $\level211 \leftrightarrow \level322 \leftrightarrow \level411$ and argued that a bosenova is also not likely in this case.
For $\alpha \gtrsim 0.3$, higher SR levels become important and solving the related system of rate equations becomes increasingly challenging.
Since the focus of the present work is to improve the statistical framework, we leave such computations for future work.

In summary, the bosenova scenario appears to be excluded for $\alpha \lesssim 0.2$.
In this regime, we may use the modified rates based on the results from \citet{2011.11646}.
For larger $\alpha$, the modified equilibrium and bosenova rates become less robust, introducing a systematic uncertainty from the assumptions underlying the related computations.
In order to compare the possible impact of the two scenarios on the ULB constraints, our results in \cref{fig:main_results} will use the SR rates as computed in the previous sections for all values of~$\alpha$.
We stress again that the rates computed in the previous sections are only robust for values of $\alpha \lesssim 0.2$ and that the bosenova scenario is strongly disfavoured in that regime.

\updated{In contrast to the above, \citet{1203.5070} performed simulations of the full Klein-Gordon equation in 3+1 dimensions on a fixed Kerr spacetime with a cosine, i.e.\ axion-like potential, for the scalar field. This method differs from the rate equation methods, and treats axion self-interactions beyond the quartic approximation. \citet{1203.5070} provided evidence that a bosenova occurs, rather than saturation. \citet{Aurrekoetxea:2024cqd} performed full numerical relativity simulations including back-reaction for binary black holes with a \emph{complex} self-interacting scalar field, and also found evidence of a bosenova.}

\updated{Finally, as already noted, BHSR for scalars has not been studied in fully non-linear $(3+1)$-dimensional numerical relativity due to the challenging timescales involved. Given this fact, and the presence of historic disagreement on the bosenova issue, we leave open the possibility of bosenova and equilibrium and present results for both in our analysis. Comparing both bosenova and equilibrium models demonstrates how our method is easily adaptable to changes in physics assumptions and allows easy comparison of the effect of different theoretical models on ULB constraints.}

\subsection{Regge trajectories}\label{sec:regge_trajectories}

As visualised by \citet[Fig.~2]{1411.2263}, the BH evolution will be dominated by the fastest SR rate for which the SR condition in \cref{eq:sr_cond} is met.
This is typically the \level211 level but, as the BH evolves and its mass and spin decrease, the SR condition will eventually cease to be fulfilled.
Once this happens, ULBs from higher levels can still drain the BH's mass and spin -- as long as they meet the SR condition and the SR rate is fast enough to make the change observable.
Eventually, there will be no such superradiant levels left, and the existence of ULBs makes a striking prediction: any given BH should not have a spin higher than a certain critical value $\astar^\text{crit}$, which depends on $M$, \tauBH, \ma, and \fax.
The line $\astar^\text{crit}(M,\tauBH,\ma,\fax)$ is referred to as `Regge trajectory'.

If we knew the true values of $(M,\astar,\tauBH)$, we could exclude the existence of any ULBs whose parameters predict $\astar^\text{crit} < \astar$.
Black holes on or below Regge trajectories might simply have been `born' with a lower spin, and this lack of knowledge about the BH's history means that they are compatible with the existence of the corresponding ULBs.
In reality, we can only estimate the BH parameters, and thus only produce a statistical estimate for the exclusion probability of a given ULB (see \cref{sec:framework}).

\section{Black hole data}\label{sec:bh_data}

We select two BH data sets as examples to demonstrate the methodology proposed in this work.
One BH is the high-mass eclipsing X-ray binary (XRB) \abhname, which has frequently been studied in the SR literature.
The other one is the SMBH \smbhname, which possesses well-defined, independent mass and spin posteriors.
It is also the first time that this SMBH has been used to infer ULB constraints.
The spin of the XRB was inferred via the thermal fit continuum method, while the spin of the SMBH was inferred from X-ray reflection spectroscopy.
We discuss these two techniques in \cref{sec:xray_binaries,sec:smbh}, respectively, and refer the reader to \citet{2011.04792} and \citet{2011.08948} for detailed, recent reviews.

\subsection{BH timescales}\label{sec:timescales}

Black holes can acquire angular momentum (be `spun up') due to accretion.
The characteristic timescale for a BH to acquire a given mass through accretion at the Eddington limit is given by the Salpeter time, $\tauSalpeter \approx \SI{4.5e7}{\yr}$, assuming a canonical radiative efficiency of 10\%.
The Salpeter time is also the appropriate characteristic timescale for BH spin-up from accretion, and thus often one simply chooses $\tauBH = \tauSalpeter$.
Although super-Eddington accretion is possible, this approach is conservative.
X-ray binaries accrete at sub-Eddington luminosities even in outburst \citep{Bahramian2023}, with only the rarer sub-population of ultraluminous X-ray sources \citep{Bachetti2014,King2023} producing super-Eddington luminosities.

Indeed, the high spin measurements in XRBs present a puzzle: it is impossible for the XRB to acquire near-maximal spin via accretion over its lifetime, thus requiring high `natal' spin \citep[for \abhname, see discussion by][]{0803.1834}.
For active galactic nuclei (AGNs), the vast majority of radiatively efficient systems accrete with Eddington fractions of $\mtx{\lambda}{Edd} \equiv \mtx{L}{bol} / \mtx{L}{Edd} \sim \numrange{0.01}{1}$ \citep{Suh2015,Temple2023}, where $\mtx{L}{bol}$ and $\mtx{L}{Edd}$ are the bolometric and Eddington luminosities, respectively.
Indeed, super-Eddington accretion is generally found to be rare in both observational \citep{Schulze2010,Zhang2024} and theoretical/simulation \citep{bustamante2019,Shirakata2019,Sala2023} studies, especially at low redshift \citep[however, see also][]{valiante-2017,Farrah2022,liu-2021,bennett-2024}.

An approximate current Eddington fraction can be calculated from the estimated bolometric luminosities and BH masses for each object, with reports of $\mtx{\lambda}{Edd} \approx 0.1$ for \abhname \citep{0803.1834} and $\mtx{\lambda}{Edd} \approx 0.2$ for \smbhname\ \citep{2009.08463}.
However, accreting BHs are variable sources, so to accurately estimate \tauBH for individual sources would require detailed knowledge of the accretion history over tens of \si{\mega\yr}.
\updated{For the SMBH case (\smbhname)}, we assume here that \updated{$\tauBH = \tauSalpeter / \mtx{\bar{\lambda}}{Edd}$}, where $\mtx{\bar{\lambda}}{Edd}$ can be thought of as an appropriately weighted time-averaged Eddington ratio.
We set $\mtx{\bar{\lambda}}{Edd} = 0.1$, which is likely to be fairly conservative, but note the inherent astrophysical uncertainty associated with this value. \updated{For the stellar mass BH M33 X-7, we instead use the system age (see section~\ref{sec:m33-x7-post})}.
If \updated{$\tauSR = \Gamma^{-1} < \tauBH/\ln N_\Delta$}, then SR dominates the evolution of the BH and \astar will decrease.
Observing a BH in a stable configuration for which we can determine \astar and \tauBH then allows us to exclude ULBs models that would predict $\tauSR < \tauBH$.

For completeness, note that the longest relevant timescale is the age of the observable Universe, which is of order the Hubble time $\mtx{\tau}{H} = 1/H_0 = \SI{14.5}{\giga\yr}$ \citep{1807.06209}.
Using $\mtx{\tau}{H}$ would result in the strongest possible limits, and it is thus a useful timescale for investigating the addressable ULB parameter space.
While the `age' of a number of SMBHs -- or, perhaps more accurately, the time since they accumulated most of their mass -- could be of order $\mtx{\tau}{H}$, as also suggested by studies with the \textit{James Webb Space Telescope} \citep[e.g.][]{2212.04568,2305.12492}, the value of \tauBH is rather set by accretion and other processes in the BH evolution, as discussed above.

\subsection{X-ray binary BHs and \abhname}\label{sec:xray_binaries}

X-ray binaries consist of a stellar mass BH accreting matter from a companion star.
XRBs are classified as low-mass (LMXB) or high-mass (HMXB), depending on whether the companion is lower or higher mass than the BH, respectively.
The accretion process emits electromagnetic radiation, which can be detected using X-ray telescopes.
Analysing the data from such measurements yields estimates for $(M,\astar)$ with, in some cases, relatively high statistical precision but potentially large systematic uncertainties due to the modelling of the system.

\subsubsection{Mass estimates in stellar-mass BHs}

As reviewed by \cite{casares2014}, the most robust method to estimate BH masses in XRBs obtains a \emph{dynamical mass measurement}, using Kepler's laws of motion and the presence of a stellar companion.
This method is more direct and robust than the methods used for the majority of BHs because the dynamical impact of the BH on its companion can be measured.
From spectroscopy, radial velocities of the emission or absorption lines from the companion star can be obtained, which allows a measurement of the orbital period $\mtx{P}{orb}$ and the radial velocity semi-amplitude~$K_c$.
The so-called `mass function' $F(M)$ can then be defined, which explicitly relates the two radial velocity parameters and $M$,
\begin{equation}
    F(M) = \frac{K_c^3 \mtx{P}{orb}}{2\pi \GN} = \frac{M \sin^3(\theta_i)}{(1+q)^2} \, ,
\end{equation}
where $\theta_i$ is the binary inclination and $q$ is the mass ratio.
For known inclination and $q$, one can estimate $M$, whereas for unknown inclination $F(M)$ provides a firm lower limit on the mass (since $1 + q > 1$ and $\sin \theta_i \leq 1$).

Dynamical mass measurements are easier to obtain for LMXBs, because the systems typically go through outburst cycles and have long periods of very low accretion activity (quiescence).
During quiescent periods, optical spectroscopy can be used to accurately characterise the radial velocity curve of the companion star, as has been carried out for about 20~LMXBs \citep{Corral-Santana2016,chaty2022}.
In HMXBs, the accretion state phenomenology is more complex; of the over 100 known HMXBs, many of which host neutron stars, only a handful of dynamical mass measurements have been possible for BH systems.
These include famous sources such as~Cygnus X-1, and the source studied here, \abhname.
In HMXBs hosting BHs, the X-ray source is persistent and true quiescent periods do not occur.
However, as HMXB companion stars are, by definition, rather massive, they outshine the accretion disc in the optical band, making spectroscopy of the donor star possible.
There are still various systematic effects that can complicate the dynamical BH mass measurements, the main three factors being line formation in (or contamination by) the strong stellar winds of the companion, uncertainties in the donor star mass, and unknown Roche lobe filling factors \citep[see discussion by][]{casares2014}.
Fortunately, \abhname is an eclipsing system, which mitigates some of the effects mentioned above.
In addition, it would be possible in future to incorporate uncertainties on, e.g., the donor mass, Roche lobe filling factor, and inclination as additional nuisance parameters into the analysis.

\subsubsection{XRB Spin Estimates from the continuum fitting method}

In addition to the X-ray reflection method (see \cref{sec:smbh_spin}), the continuum-fitting method has widely been used to estimate BH spins in XRBs.
This relies on the influence of BH spin on the inner edge of the accretion disc, as higher (prograde) spins lead to an innermost stable circular orbit (ISCO) lying closer to the BH.
The closer the ISCO radius ($\mtx{R}{ISCO}$) is to the event horizon \citep[which depends on $M$ and \astar; cf.\ e.g.\ Eqs~(2)--(4) in][]{2011.08948}, the more energy will be released per unit accreted matter.
The maximum effective temperature of the accretion disc corresponds to that at the ISCO and is set by $\mtx{T}{eff,max}^4 \propto \mtx{\lambda}{Edd} M \mtx{R}{ISCO}^{-3}$ in the thin-disc model.
For a canonical 10\% radiative efficiency, $\mtx{T}{eff,max}$ peaks in the soft X-ray band (i.e.\ below $\SI{2}{\keV}$) for an accreting stellar BH of $M = 10\,\Msol$.
In contrast, $\mtx{T}{eff,max}$ peaks in the extreme UV band for an accreting SMBH with $M = 10^6\,\Msol$ for the same canonical radiative efficiency.
The interpretation of observed X-ray spectra of XRBs accreting at moderate rates relative to the Eddington limit ($\mtx{\lambda}{Edd} = 0.01-0.3$), paired with suitable accretion disc models, has provided spin constraints for around 10 XRBs.
We refer to \citet[][]{steiner2009,McClintock2014,2011.08948} for relevant reviews.

Generally, the accretion disc model assumed in continuum-fitting spin studies is the geometrically-thin, optically-thick, steady-state accretion disc model due to \citet{novikov1973}.
This model is an extension to the also well-known Shakura--Sunyaev alpha thin-disc model \citep{shakura1973} to the relativistic regime.
In the Novikov--Thorne model, mass loss due to winds in the inner disc is assumed to be negligible, and, as in other thin-disc models, the disc acquires non-molecular viscosity due to internal stresses and heat is dissipated locally.
These two conditions have fairly sound theoretical and empirical motivation, given that thin-disc models can describe the putative disc component of the observed X-ray spectrum in XRBs during soft, thermally dominated accretion states.
Furthermore, fully relativistic continuum-fitting models generally adopt a zero-torque boundary condition at the ISCO.

Continuum-fitting models predict local thermal spectra by combining the predicted thermal disc spectrum -- subject to an accretion disc model -- with the temperature-dependent colour correction factor.
Although the temperature dependence of the latter can be inferred from sophisticated radiative transfer calculations \citep{davis2005}, in practise, some continuum-fitting models implement an approximate temperature-dependent colour correction factor, such as that described by \cite{done2012}.

Spin estimates in XRBs from continuum fitting have been achieved with high statistical precision.
However, the systematic errors emerging from the assumptions and approximations made by these models must be accounted for.
First, as mentioned above, the thin-disc solution is only expected to be appropriate for thermally-dominated discs for bolometric luminosities $\mtx{L}{bol} \lesssim 0.3 \, \mtx{L}{Edd}$ \citep[][]{kulkarni2011}.
Second, even in this moderate luminosity regime, radiation could be emitted from within the ISCO, which is expected to induce an uncertainty in the recovered spin from continuum fitting of $\sigma_{\astar} = 0.05$ \citep{zhu2012}.
Third, if the aforementioned zero-torque boundary condition at the ISCO is not satisfied, e.g.\ due to magnetic fields threading the inner disc, further systematic errors of $\sigma_{\astar} \leq 0.1$ would be induced \citep{Shafee2008-a,Shafee2008-b}.

Moreover, astrophysical parameters with inherent uncertainties -- such as the inclination to the angular momentum vector of the binary, the BH mass, and the distance and accretion geometry of the XRB system -- are also expected to contribute to the overall systematic uncertainty on the inferred BH spin.
These nuances have led to variations in the analysis pipelines employed by different research groups amongst different data sets, potentially leading to further systematics from modelling and different statistical methods.
Nevertheless, a consistent analysis of multiple XRBs would in principle be possible, thanks to publicly available tools such as the \code{Xspec} spectral-fitting package \citep{1996ASPC..101...17A}.

\subsubsection{\abhname posteriors}
\label{sec:m33-x7-post}
In this work, we use the spin posterior of the XRB system \abhname of \citet{0803.1834}, inferred using a fully relativistic continuum-fitting model.
The analysis in \citet{0803.1834} was applied to a set of \textit{Chandra} and \textit{XMM-Newton} observations of the binary in the thermally-dominated state and applied to epochs where $\mtx{L}{bol} < 0.3 \, \mtx{L}{Edd}$.
Based on previously determined system inclination and BH mass values, \citet{0803.1834} constrained the BH spin value to $\astar = \num{0.84(5)}$.
The authors noted that the assumptions made by their fully relativistic continuum-fitting model, added to the assumption that the angular momentum vector of the binary is approximately aligned with the BH spin vector, are expected to induce additional systematic uncertainties.

\begin{figure}
    \centering
    \includegraphics[width=3.38in]{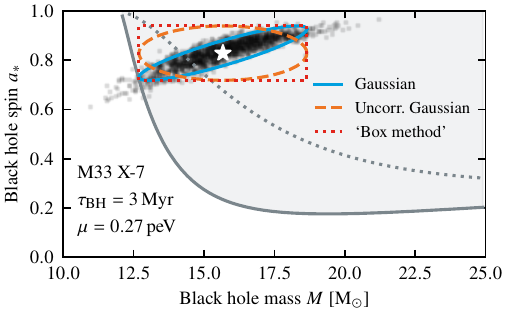}
    \caption{The sampled BH mass and spin distribution (black points) of \abhname and Regge trajectories of the non-interacting (grey lines and shaded region) and self-interacting \level211 level (dotted, grey line; equilibrium regime, $\ifax = \SI{2e-15}{\GeV^{-1}}$), $\ma = \SI{0.27e-12}{\eV}$, and $\tauBH = \SI{3e6}{\yr}$. The full distribution, whose mean is marked with a white star, is compared to its $2\sigma$ error bars for an uncorrelated (dashed, orange line) and full (blue line) Gaussian, and a `$2\sigma$ box' (dotted, red line).\label{fig:distr_m33x7}}
\end{figure}

In \cref{fig:distr_m33x7}, we show the parameter sampling distributions from bootstrapping (black dots) for the BH system \abhname, which we obtained by digitising Fig.~3 of \citet{0803.1834} \citep[see also][]{0710.3165}.
Clearly, the $2\sigma$ contour of an uncorrelated Gaussian (dashed, orange line) does not provide a good description of the distribution, and including correlations (blue line) is more appropriate.
We also show the `box' defined by the uncorrelated $2\sigma$ errors, which captures all possible underlying correlations at the price of discarding the information contained therein.
We will discuss the relative importance of these effects on the ULB limits in \cref{sec:comparison}.

Note that, in other cases, the resulting $(M,\astar)$ distribution may be highly non-Gaussian.
For instance, the somewhat uniformly distributed posterior distribution of GRS~1124-683 \citep{1601.00615} provides an extreme example.

\Cref{fig:distr_m33x7} also shows the Regge trajectory (solid, grey line and shading; cf.\ \cref{sec:regge_trajectories}) for the non-interacting \level211 level and an ULB mass of $\ma = \SI{0.15e-12}{\eV}$.
Since the estimated age $\tauAge = \SI{3e6}{\yr}$ of \abhname \citep{1106.3690} is below $0.1\,\tauSalpeter$, we choose $\tauBH = \tauAge$.
To illustrate the effect of self-interactions, in this case assuming the equilibrium regime and $\ifax = \SI{4e-16}{\GeV^{-1}}$, we also include the modified Regge trajectory as a dotted, grey line.
We can exclude ULBs if most of the $(M,\astar)$ samples can be found above the Regge trajectory.

\subsubsection{Effects of the companion star}

\updated{The companion of \abhname has a mass of $M_2 \approx 70\,\Msol$ \citep{1010.4742} and an orbital period of $T = \SI{3.45}{\day}$ \citep{astro-ph/0603698}. Using the equations from Appendix~I of \cite{2011.11646}, we verified that the companion does not have a significant effect on the exclusion regions for the ULB parameters.}~\footnote{This conclusion is consistent with those reached by \cite{Witte:2024drg}.}
\updated{For other systems, however, the companion might play a more important role. In such cases, our proposed statistical framework can include and propagate the related uncertainties on the companion parameters, which would be particularly relevant for the large uncertainty of about 10\% on $M_2$ \citep{1010.4742} (see \cref{sec:conclusions} for other such nuisance parameters).}

\subsection{Supermassive BHs and \smbhname}\label{sec:smbh}

Supermassive black holes are believed to reside at the centres of all galaxies, and a compelling way to study them observationally is by probing the activity of their associated active galactic nuclei (AGN).
AGNs are nuclear regions located at the centre of about $\numrange{1}{10}\%$ of galaxies.
Their integrated luminosity exceeds the stellar luminosity of their host galaxies as a consequence of the accretion onto the SMBH from the surrounding accretion disc.
Given the high-quality observations needed, obtaining spin and mass estimates for SMBHs is rather challenging.
Here, we summarise how relativistic X-ray reflection and infrared interferometry can respectively be used to constrain the spin and mass of SMBHs in AGNs.

\subsubsection{SMBH mass estimates}\label{sec:smbh_mass}

The mass of {Sgr~A*}, the closest SMBH to Earth, can be measured dynamically from the motion of stars within the BH gravitational potential \citep{Ghez2008,Genzel2010}.
However, in more distant AGNs, the BH's sphere of influence cannot be resolved and its mass is often estimated from broad emission lines: under the assumption that the broad-line region (BLR) gas is virialised, the width of the line, $\Delta v$ is related to the BH mass through the relation $M = \colfac \, \mtx{R}{BLR} (\Delta v)^2 / \GN$, where \colfac is the so-called virial factor, $\colfac \sim \order{1}$, that accounts for BLR geometry and projection effects.
Using this virial estimator requires an estimate for the line formation radius $\mtx{R}{BLR}$, which ideally would be estimated from reverberation lags \citep{kaspi2000,kaspi2005}.
If $\mtx{R}{BLR}$ is not known, then it is usually estimated from the luminosity-size relation, $\mtx{R}{BLR} \propto L^{1/2}$ -- where $L$ is an appropriately chosen proxy for the bolometric or ionising luminosity -- calibrated with the low-redshift empirical relation from reverberation mapped AGNs \citep[e.g.][]{bentz2013}.
There are numerous other ways of estimating BH mass, which we cannot describe exhaustively, each with their own characteristic uncertainties and systematics; notable methods include the use of scaling relations between $M$ and the stellar velocity dispersion \citep{ferrarese2000,gebhardt2000} or bulge mass \citep{magorrian1998}, and the observation of \ce{H_2O} megamasers \citep{greene2010}.

Recently, the GRAVITY interferometric beam-combiner \citep{1705.02345} on the \textit{Very Large Telescope} has made it possible to estimate $\mtx{R}{BLR}$ and model the BLR geometry from the differential phases in infra-red interferometry.
These measurements are extremely useful for BH masses and have allowed more accurate estimates for a number of luminous AGNs \citep{gravity_quasar_2018,gravity_3783_2021}.
Additionally, while subject to various assumptions, the modelling provides a well-defined posterior probability distribution for $M$ for our analysis.
One such object observed is \smbhname, which is our chosen SMBH case study (see \cref{sec:smbh_posterior}).
Our use of the mass posterior from GRAVITY observations mitigates some, but not all, of the systematic uncertainties associated with BH mass estimates.

\subsubsection{SMBH Spin estimates from X-ray reflection spectroscopy}\label{sec:smbh_spin}

In this work, we adopt the BH spin posterior distribution for \smbhname inferred by \cite{walton2020}, who used the X-ray reflection method.
In the absence of Compton-thick absorption and of spectral degeneracies induced by warm absorbers and ultra-fast outflows, the detailed modelling of reflected features in X-ray spectrum of AGNs provides a powerful pathway to probing the strong gravity regime in their central SMBHs.
We note that several alternative observational techniques have been used to generally probe spins of BHs accreting at lower rates or hosting strong radio-jets \citep{unal_on_2020,ricarte_ngEHT_2023,daly_robust_2022,wen_mass_2021,tamburini_measurement_2020}.
We refer to \citet{davoudiasl_ultralight_2019} for a discussion on BHSR from the spin estimate of {M87*} from the interpretation of \textit{Event Horizon Telescope} data; and to \citet{2011.06010} for joint constraints on the ultralight boson mass and spin distribution at the time of binary BH formation from the Gravitational-Wave Transient Catalog (GWTC-2).

The disc reflection spectrum arises from the reprocessing of the primary X-ray continuum from the inner and outer regions of the ionised accretion disc, as well as from distant neutral and non-relativistic material surrounding the accretion disc.
The primary continuum is thought to originate from the Compton up-scattering of seed photons in the innermost regions of the disc by the electrons in the hot X-ray corona.
A fraction of the direct coronal emission towards the observer will back-scatter on the disc and be re-radiated away, with the non-thermal emission of coronal electrons governing the high-energy cutoff tail of the broadband spectral energy distribution (usually in the \SIrange{100}{300}{\keV} range) in type-1 AGNs.

Amongst other features in the X-ray reflection spectrum, the most prominent one is the \FeKalph line, as iron is the most abundant element in the disc and has a high fluorescent yield~\citep{george:91a}.
Generally, the \FeKalph line receives contributions from reprocessing from the inner regions of the disc, as well as cold, non-relativistic matter surrounding the immediate vicinity of the AGN.
A narrow \FeKalph reflection feature (centred at \SI{6.4}{\keV} in the rest frame) often arises due to the reflection from cold matter surrounding the outer accretion disc.
However, the most important feature for BH spin inference is the broad \FeKalph feature, centred around \SIrange{6.4}{6.97}{\keV} rest-frame energy, which likely indicates the presence of reprocessed emission from the innermost regions of the accretion disc.

Given that this reprocessing is thought to take place in close proximity to the BH event horizon, the reflected emission is subject to standard Doppler and general relativity effects in the disc, with the latter including light bending and gravitational redshift.
The spectral imprints emanating from these effects, especially a distinct red wing to the line, will be more dramatic and noticeable the closer the emission takes place to the event horizon~\citep{fabian_x-ray_1989}.
In addition, relativistic reflection from the innermost regions of the disc will also induce a forest of reflection features down to the soft X-ray band (below \SI{2}{\keV}).
Finally, another characteristic feature of the disc reflection spectrum in AGNs is the Compton (reflection) hump, which arises in the range of \SIrange{20}{40}{\keV} due to photo-electric absorption of low-energy photons and the multiple reprocessing of high-energy (coronal) electrons \citep{zdziarski:90a}.

Current relativistic X-ray reflection models can provide a powerful probe of the BH spin if the inner accretion disc extends down to $\mtx{R}{ISCO}$.
Under this condition and assuming an accretion disc model, these models can then be used to make predictions of reflected spectra parameterised by \astar, given assumptions about:
\begin{itemize}
    \item the geometry of the corona, 
    \item the disc density at the mid-plane and/or disc ionisation state, 
    \item and the metric of spacetime around the central BH.
\end{itemize}

State-of-the-art X-ray reflection models \citep[such as the widely used \texttt{relxill} family of models, described by][]{garcia_improved_2014,dauser_role_2014,dauser_normalizing_2016} have provided spin constraints for about 20 stellar-mass BHs and 50 SMBHs \citep{2011.08948}.
The simplifications made by these reflection models, and their possible translation into systematic biases on the inferred BH spin were thoroughly reviewed by \citet[Sec~3.2]{2011.08948} and \citet[Secs~3 and~4]{2011.04792}.
We also note that, although the majority of current spin estimates drawn from X-ray reflection spectroscopy rely on the detection and interpretation of the broadened \FeKalph line, several SMBH spins in AGNs have been inferred via the reflection interpretation of the soft excess -- notably, in Narrow-line Seyfert 1 galaxies \citep{jiang_2019,mallick_high_disc_2022}.

\subsubsection{\smbhname\ posterior samples}
\label{sec:smbh_posterior}

In this work, we make use of spin posteriors from a specific analysis of \smbhname by \citet{walton2020}, who analysed the broadband X-ray data from the \textit{Swift}, \textit{XMM-Newton}, and \textit{NuSTAR} telescopes.
As is commonly the case, this AGN is found to have a complex X-ray spectrum with signatures of X-ray reflection from the inner accretion disc as well as both neutral and ionised absorption by more distant circumnuclear matter along the line of sight.
\citet{walton2020} construct a 23-parameter spectral model to describe the primary continuum reflection from a BH accretion disc, which then passes through multiple layers of absorption \citep[see Table 4 of][]{walton2020}.
Using that model, and the multi-satellite dataset, the authors computed the the posterior distribution of the parameters with the Goodman--Weare algorithm.

Marginalising over all other parameters of this global model, the inferred black hole spin is $\astar = 0.94^{+0.02}_{-0.07}$.
The principal `nuisance' parameters that have bearing on the spin are the inclination of the accretion disc, the iron abundance of the disc, the ionisation state of the surface layers of the disc, and the geometry of the X-ray-emitting corona (characterised as a `lamppost' height, also known as the coronal geometry).
For typical AGN parameters, the inclination of the disc is readily determined by the high-energy `edge' of the iron line profile, resulting in little degeneracy with the inferred spin (that results from the shape of the low-energy `tail' of the iron line).
The iron abundance principally influences the relative strength of the iron line and the Compton reflection hump with secondary impacts on the inferred spin, but is well constrained in datasets that cover both features (such as is the case here). 
The ionisation state of the accretion disc photosphere strongly influences the shape of the low-energy ($E < \SI{4}{\keV}$) reflection spectrum and shifts the energy of the iron line but, again, is well constrained by datasets that cover the \SIrange{0.5}{50}{\keV} band.
It is the lamppost height that has the greatest degeneracy with the inferred spin; this is included in the quoted errors on the spin measurement \cite[see Figs~A1--A2 of][for the associated corner plots]{walton2020}.

Our mass posteriors come from the BLR modelling in the analysis of interferometric data by the \citet{1705.02345,2009.08463}.
This mass estimate is consistent with the estimate from timing by \citet{walton2020}.

\begin{figure}
    \centering
    \includegraphics[width=3.38in]{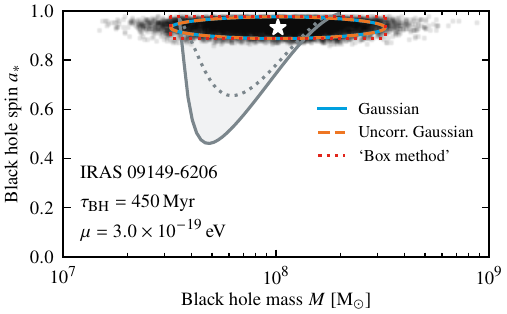}
    \caption{The sampled BH mass and spin distribution (black points) of \smbhname and Regge trajectories of the non-interacting (grey lines and shaded region) and self-interacting \level211 level (dotted, grey line; equilibrium regime, $\ifax = \SI{3e-16}{\GeV^{-1}}$), $\ma = \SI{3e-19}{\eV}$, and $\tauBH = \SI{4.5e8}{\yr}$. The full distribution, whose median is marked with a white star, is compared to its $2\sigma$ error bars for an uncorrelated (dashed, orange line) and full (blue line) Gaussian, and a `$2\sigma$ box' (dotted, black lines).\label{fig:distr_iras09149_6206}}
\end{figure}

We show the \smbhname posterior samples for $M$ \citep{1705.02345,2009.08463} and \astar \citep{walton2020} in \cref{fig:distr_iras09149_6206}, assuming a BH timescale of \updated{$\tauBH = \tauSalpeter/0.1 = \SI{4.5e8}{\yr}$}.
Note that, due to the two independent measurements for $M$ and \astar, \emph{there are no correlations} between the two parameters in this case.
As for \abhname, we again include the Regge trajectories for non-interacting (solid, grey line) and self-interacting (dotted, grey line) ULBs.

Since \smbhname has a spin close to the theoretical maximum value, the `$2\sigma$~region' of the \astar~distribution surpasses that value.
Instead of a regular Gaussian, one could use a truncated Gaussian to implement this physical upper limit on \astar.

We also checked that, as is presumably the case for many SMBHs, the $M$~distribution is better described by a Gaussian in $\ln(M/\Msol)$ than in~$M$.
The mean value of $\ln(M/\Msol)$ corresponds to the median for the log-normal distribution of~$M$, and Gaussian error propagation, $\sigma_M/M \approx |\sigma_{\ln(M/\Msol)}|$, is not necessarily a good approximation.
Converting quoted errors on $\ln(M/\Msol)$ into an error on~$M$ is thus potentially problematic -- even more so when considering the `$2\sigma$~region'.
We will see the effect of this in \cref{sec:comparison}.

\subsection{Gravitational-wave binary BH mergers}\label{sec:gw_bhs}

Gravitational-wave (GW) data can be used to estimate the mass and spin of binary black hole (BBH) mergers.
In principle, these data can thus be used to constrain ULB properties.

However, current detectors only measure a mass-weighted combination of the spins along the orbital angular momentum \citep{1805.03046}.
This complicates the determination of the individual spins at the time of the merger, while the unobservable pre-merger history also introduces additional challenges \citep{1908.02312,2011.11646}.
These caveats apply to ULB constraints from BHB mergers based on the inferred pre-merger spins alone, as e.g.\ presented by \citet{1805.02016,1911.07862,2009.07206}.

To ameliorate these issues, \citet{1908.02312,2011.06010,2201.11338} propose using a Bayesian hierarchical analysis to account for the pre-merger history using hyperprior distributions.
\citet{2201.11338} in particular investigate the influence of prior choices and consider an average of possible BBH lifetimes in the range of \SIrange{e6}{e10}{\yr}.
Since the constraints are mainly driven by a few of the observed BBH systems, the authors conclude that more data is needed for more robust conclusions once a broad range of pre-merger scenarios is included in the analysis.
We thus leave the including of GW data from BBH mergers for future work.

\section{Statistical framework}\label{sec:framework}

With the BHSR rates from \cref{sec:equilibrium,sec:bosenovae}, and the posterior distributions for the BH parameters from \cref{sec:xray_binaries,sec:smbh}, we can now introduce our procedure for computing exclusion regions for ULB models.

As discussed in \cref{sec:regge_trajectories}, we can exclude ULB models\footnote{Thanks to \cref{eq:qcd_axion_mass}, QCD axion models effectively only have one free parameter (either \ma or \ifax), in addition to the (well-constrained) topological susceptibility \chiTop as a nuisance parameter.} $\vc{\alpha} = (\ma,\ifax)$ for which the \emph{true} values of $\vc{\beta} = (M,\astar,\tauBH,\dots)$ lie above the Regge trajectory, i.e.\ when $\astar > \astar^\text{crit}(M,\tauBH,\dots,\vc{\alpha})$.
The ellipsis in $\vc{\beta}$ is to indicate that there may be more parameters of the BH environment that have an influence on the Regge slope.
For this work, however, we will not consider any such parameters and assume \tauBH to be a fixed value.
Also note that any other BH nuisance parameters, such as the inclination angle~$i$ or virial factor~\colfac, can already be marginalised out in the original BH analysis.

In terms of probabilities, we can write the condition on \astar as a Heaviside function, $p(\astar | M,\vc{\alpha}) = \Theta(\astar^\text{crit} - \astar)$, i.e.\ $p(\astar | M,\vc{\alpha}) = 0$ if the BH lies above the Regge slope, and otherwise $p(\astar | M,\vc{\alpha}) = 1$.
In particular, the BH's parameters only depend on the ULB parameters when SR is active and the the condition on \astar applies, such that the conditional probability $p(\vc{\beta}|\vc{\alpha})$ factorises as $p(\vc{\beta}|\vc{\alpha}) = p(\vc{\beta}) \, p(\astar | M,\vc{\alpha})$.

While we do not know the true values of $\vc{\beta}$, we can obtain a sampling distribution from the BH data~$\data$ to compute the posterior $p(\vc{\alpha}|\data)$.
Using the law of conditional probabilities and Bayes' theorem, we find that
\begin{align}
    p(\vc{\alpha} | \data)
    &= \int \! p(\vc{\alpha}, \vc{\beta} | \data) \, \dd \vc{\beta} 
    = \int \! p(\vc{\beta} | \data) \, p(\vc{\alpha} | \vc{\beta}, \cancel{\data}) \, \dd \vc{\beta} \\
    &= \int \! p(\vc{\beta} | \data) \, \frac{p(\vc{\beta} | \vc{\alpha}) \, p(\vc{\alpha})}{p(\vc{\beta})} \, \dd \vc{\beta} \\
    &= p(\vc{\alpha}) \int \! p(\vc{\beta} | \data) \, p(\astar | M, \vc{\alpha}) \, \dd \vc{\beta} \label{eq:posterior} \\ 
    &\approx \frac{p(\vc{\alpha})}{N} \sum_{i=1}^{N} p(\astar^i | M^i, \vc{\alpha}) \, ,\label{eq:posterior_sampling}
\end{align}
where we use a slash to indicate the removal of the redundant dependence on the data~$\data$ and where, in the last step, we assume that the BH analysis provides $N$ equally-weighted samples $\vc{\beta}^i = (M^i,\astar^i)$ from the posterior $p(\vc{\beta} | \data)$, which we can use to compute the integral in \cref{eq:posterior} via Monte Carlo integration.

Since $p(\vc{\alpha} | \data) \propto p(\vc{\alpha}) \, p(\data | \vc{\alpha})$, we may re-interpret the sum in \cref{eq:posterior_sampling} as a marginal likelihood.
Furthermore, the computation of $p(\astar | M,\vc{\alpha})$ requires a computation of the BHSR rates and assumptions about the BH evolution.
As we will discuss in \cref{sec:limitations}, both our computations and assumptions could be refined further.
Even then, the framework presented here is general enough to allow for such modifications without having to change to the framework itself.

\section{Results and Discussion}\label{sec:results}

\begin{figure*}
    \centering
    \includegraphics[width=6.76in]{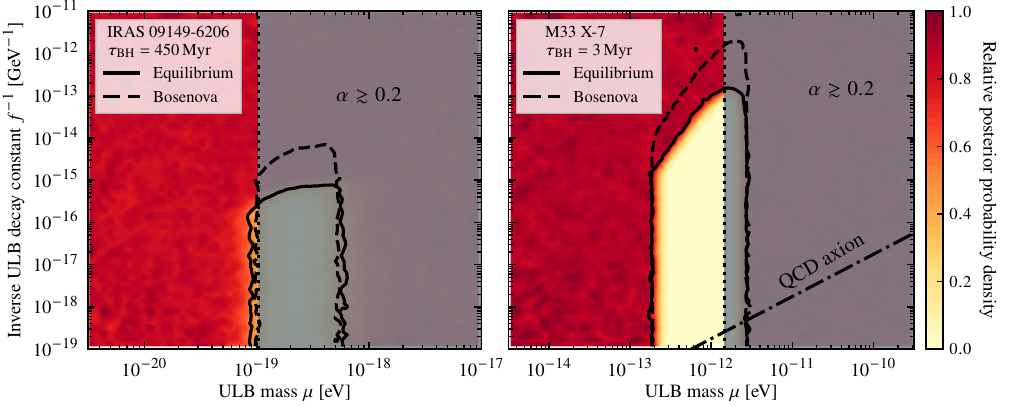}
    \caption{Normalised posterior probability density distributions for $(\ma,\ifax)$ for \smbhname (\textit{left}) and \abhname (\textit{right}). We show the 95\% credible regions at highest posterior density for both the equilibrium (solid line) and bosenova (dashed lines) scenarios and, in the right panel, the QCD axion model line (dashed-dotted line) predicted by \cref{eq:qcd_axion_mass}). Note that we highlight (grey shading) the \ma region where more than 5\% of BH mass samples imply $\alpha > 0.2$ (dotted line) and, meaning that the computation of the superradiance rate involves large theoretical uncertainties and may not be valid.
    \label{fig:main_results}}
\end{figure*}

Our main results can be found in \cref{fig:main_results}, where we show the posterior distribution for $(\ma,\ifax)$ for \smbhname (left panel) and \abhname (right panel).

We assume log-uniform prior distributions for \ma and \ifax, limited by the ranges shown in \cref{fig:main_results}.
The ranges approximately correspond to the relevant (observable) parameter space, except that we restrict $\fax < \Mpl$.
The latter is expected in string theory, and more generally if the effective field theory is free of corrections from quantum gravity \citep{Arkani-Hamed:2006emk}.

We use \code{emcee}~\citep{1202.3665} with 48~walkers to generate, after a burn-in run, \num{60000} samples per chain i.e.\ a total of around \num{e7} samples (total runtime around 600\,CPUh).
Finally, we compute the 95\% credible regions (CRs) at highest posterior density (HPD) for the ULB parameters.

We can see from \cref{fig:main_results} that the equilibrium regime and the bosenova condition lead to similar disfavoured regions of parameter space in \ma.
In particular, the lower \ma cutoff is set by the condition to achieve a significant cloud occupation while, at larger \ma, the limits are cut off by the SR condition.
It has been noted before, and we will see this explicitly in \cref{sec:comparison}, that the range of excluded~\ma may be extended by including higher occupation levels with $n > 2$.
However, here we only show results considering the \level211 since our computed SR rates are most reliable for this case.
To further emphasise this, we also show the \ma value for which we mostly, i.e.\ for at least 95\% of the $M$ samples for a given BH, obtain $\alpha < 0.2$.
The ULB constraints are most robust in this regime where, moreover, the occurrence of bosenovae is strongly disfavoured (see \cref{sec:bn_vs_eq}).

From the relative posterior densities, it appears that the constraints on \smbhname are weaker than for \abhname, which we will explore in more detail in \cref{sec:comparison}.
This highlights the relevance of a precise determination of SMBH masses to obtain meaningful constraints.
While only BHs with a high spin can give strong limits on ULB parameters, the connection between $M$ and \ma via $\alpha = M \ma$ implies that we can only locate these constraints if $M$ is measured precisely enough.
Large uncertainties on $M$ prevent us from doing so and, although it is clear that such a BH would give strong constraints on ULBs, an improvement in the BH mass measurement is required to tell us about the corresponding \ma range.
The primary and secondary criteria for suitable BH candidates for obtaining ULB constraints should thus be a high \astar and precisely determined~$M$, respectively.

Note that the adopted priors implicitly assume the existence of ULBs within their possible parameter range, and that we can always define credible regions of the posterior distribution to exclude part of the ULB parameter space.
Such constraints are only meaningful if the prior is sensible -- i.e.\ in particular \emph{if such ULBs exist}.
In order to assess the probability of ULBs' existence, one has to perform a model comparison, e.g.\ using Bayes' factors.
Here, we are interested in setting limits, and thus do not perform this additional step.

More generally, the choice of prior on $(\ma,\ifax)$ will necessarily affect the exclusion regions.
\citet{1810.07192} analysed this in more detail for axion models and found that the prior dependence can be very strong.
Still, at least with regards to changing the range of the log-uniform priors, we do not expected extreme deviations as for e.g.\ the choice between log-uniform and uniform priors.
In particular, when combining data from multiple BHs, the overall prior range will increase when considering the `sensitive \ma region' for all BHs, making a joint fit increasingly conservative.

We also want to stress that the marginal distributions on \ma and \ifax will not be very constraining:\footnote{This is different for QCD~axions, due to the connection between \ma and \fax from \cref{eq:qcd_axion_mass}.
Choosing a Gaussian prior on \chiTop, we can compute the posterior distribution of $(\ifax,\chiTop)$ for \abhname, which constrains the relevant portion of parameter space (cf.\ \cref{fig:main_results}).
In both the equilibrium and bosenova scenarios, we find that $\SI{0.6}{\pico\eV} < \ma < \SI{3}{\pico\eV}$ for the 95\% credible interval, where the lower end of the interval is effectively set by the prior ($\fax < \Mpl$).} there is practically no limit on \ifax alone due to the available \ma parameter space with high posterior density outside of the region related to the inferred $M$ range.
Similarly, the region of \ma that can be constrained when marginalising \ifax is much narrower than suggested by the two-dimensional exclusion region.
When considering self-interactions, it is important to show the two-dimensional contour instead of assuming that the excluded one-dimensional \ma range will correspond to the range suggested by the two-dimensional contour at low \ifax.
This would even be more so the case in a frequentist framework.

\subsection{Comparison with other works}\label{sec:comparison}

Surveying the literature, we identify two main strategies to obtain ULB constraints, which we refer to as the `box method' and ` Gaussian distance method', respectively.

\paragraph*{The `box method'.} \citet{1411.2263} applied the following checks for BH parameter estimates $(\hat{M},\astarhat)$ with associated uncertainties $(\sigma_{\hat{M}},\sigma_{\hat{a}})$ and $\tau_\text{BH} = \min(\tauAge, \SI{40}{\mega\yr})$.
A given ULB model can be excluded if at least one \state{nlm} with $l \leq 5$ (i.e.\ $n \leq 6$) and \emph{all} $M \in [\hat{M} - 2\sigma_{\hat{M}}, \hat{M} + 2\sigma_{\hat{M}}]$ and $a = \hat{a} - 2\sigma_{\hat{a}}$ fulfil the SR and bosenova conditions (see \cref{sec:bosenovae}).
Comparing to \cref{fig:distr_m33x7,fig:distr_iras09149_6206}, these conditions demand that the rectangle defined by the mass and spin measurement $\pm 2 \sigma$ in each direction is contained within the envelope of the Regge trajectory, and that bosenovae do not spoil the spindown of the BHs.
\citet{2011.11646} also used the `box method', but based on the equilibrium scenario and conditions (see \cref{sec:equilibrium}).

\paragraph*{The `Gaussian distance method'.} \citet{1805.02016,2009.07206} proposed to interpret $(\hat{M}, \astarhat)$ estimates and their errors as uncorrelated Gaussian measurements.
They then defined `effective errors' with respect to the Regge trajectory, which allowed them to compute a distance measure between the uncorrelated Gaussian distribution of the measurement and the area above the Regge trajectory.
Their goal was to compute the weighted overlap between Gaussian distribution and the Regge trajectory ``in a numerically efficient manner'', as visualised in \citet[Fig.~17]{1805.02016}.
In this sense, their underlying logic is similar to ours with the difference being their assumption about the uncorrelated Gaussian distribution and their computational scheme which will, as they pointed out, only approximately give the desired result when considering levels with $n > 2$ \citep[Appendix~B]{1805.02016}.

\begin{figure*}
    \centering
    \includegraphics[width=6.76in]{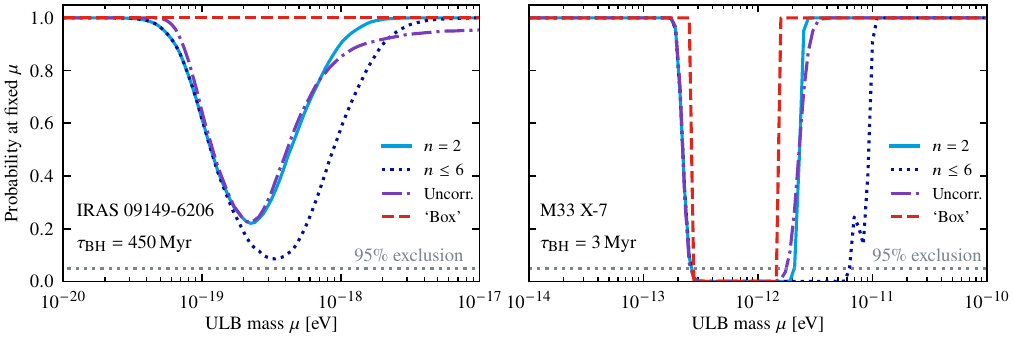}
    \caption{Comparison of exclusion methods at fixed mass \ma for non-interacting ULBs for \smbhname (\textit{left}) and \abhname (\textit{right}). We compare our approach for the full posterior distribution with $n = 2$ (solid blue lines) and $n \leq 6 $ (dotted blue lines) compared to uncorrelated Gaussians (`Uncorr.'; dashed-dotted, purple line) and the `box method'~(`Box'; dashed, red line).}
    \label{fig:comparison}
\end{figure*}

\paragraph*{Comparison with our method.} To facilitate a direct comparison with the literature in \cref{fig:comparison}, we first ignore self-interactions -- i.e.\ take $\fax \to \infty$ or, equivalently, set $\lambda = 0$ -- and compute the posterior probability at fixed values of \ma.\footnote{This technically corresponds to repeating the analysis multiple times with different `$\delta$~priors'.}
Instead of implementing the `Gaussian distance method', we will use our method, but with an uncorrelated Gaussian instead of the full posterior distributions.
This is equivalent to what \citet{1805.02016} wanted to compute.
For illustrative purposes, for the uncorrelated Gaussian, we choose estimators in $M$ for both BHs, even though the underlying distribution of the SMBH is better described by a Gaussian in $\log(M/\Msol)$.
Finally, note that the `box method' gives a binary outcome; the probability is either $P = 0$ or $P = 1$.

Consider the left panel of \cref{fig:comparison} for \smbhname.
The `box method' finds $P = 1$ for all ULB masses and, indeed, the lines for the other methods and assumptions also remain above the exclusion threshold.
However, thanks to our probabilistic interpretation, our method could yield constraints by combining the data from multiple BHs -- even when they individually would not constraint ULB parameters.
Further note that the deviations between the full and the approximated results mainly come from the approximate Gaussian error propagation, as the data themselves are uncorrelated.
\Cref{fig:comparison} also demonstrates explicitly that the inclusion of more levels can extend the bounds to higher ULB masses.
This is because the higher $n$, and thus higher $l$, allow us to fulfil the SR condition in \cref{eq:sr_cond} for larger \ma.
For these constraints, we only consider the SR rates $\Gamma_{\nlm}$ of the individual levels with no regards to the BH evolution and time required to fill these levels.

The case of \abhname is shown in the right panel of \cref{fig:comparison}.
Given the clearly visible correlations in the $(M,\astar)$~distribution in \cref{fig:distr_m33x7}, it is somewhat surprising that neglecting them appears to be an acceptable approximation.
This might only be a somewhat lucky accident for \abhname, but at least in this case the approximation seems to be inconsequential.
The inclusion of levels with $n \leq 6$ again extends the ULB mass range of the constraints.

\begin{figure}
    \centering
    \includegraphics[width=3.38in]{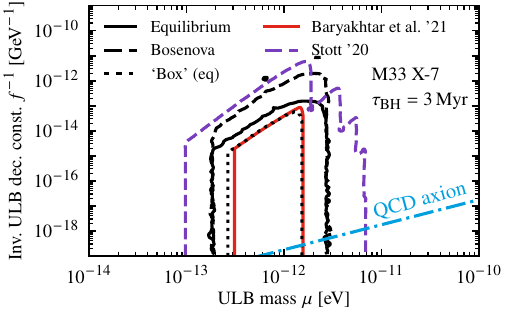}
    \caption{Comparison of exclusion contours for $(\ma,\ifax)$ with different methods and assumptions for \abhname. We compare our results from \cref{fig:main_results} to two recent works using the `box' and `Gaussian distance' methods. The QCD axion model line is shown as a dashed-dotted, blue line).}
    \label{fig:comparison_m33x7}
\end{figure}

\Cref{fig:comparison_m33x7} presents a direct comparison with ULB constraints from \abhname in the $(\ma,\ifax)$ plane.
\citet{2009.07206} used the bosenova prescription for self-interactions, a Gaussian approximation for the BH errors, and included levels up to $n = 6$ (dashed, purple line).
The result broadly agrees with our implementation (dashed, black line) of the bosenova prescription in terms of \ifax, while we observe that the inclusion of higher SR levels extends the limits from \cite{2009.07206} to higher values of \ma.
Our results disagree with \cite{2009.07206} at the low mass end by a factor of $\mathcal{O}(2)$ on \ma.
This is due to \cite{2009.07206} \citep[and the related works][]{1805.02016,2103.06812} neglecting the large $\ln(\NMax)$ factor required to extract significant spin, and the effect of the companion.
\cite{2011.11646} used the `box' statistical method, and the equilibrium model for self-interactions (solid, red line).
As mentioned before, our result (solid, black line) gives rise to a slightly wider exclusion region on \ma, due to the overly conservative nature of the box method.
Note that our implementation of the box method (dotted, black line) closely reproduces the results by \cite{2011.11646} \updated{for both $\ma$ and $\ifax$}.

Advantages of our approach include that we do not assume Gaussianity, which is not always justified -- in particular for the treatment of upper limits on \astar.
It is important to stress again that, for SMBHs, Gaussianity only appears to be justified for $\ln(M/\Msol)$ rather than $M$.
While past work also ignored potential correlations between $M$ and \astar, we do not find this to be a major issue in our examples.
Our approach also turns out to be computationally feasible with Monte Carlo integration, alleviating concerns of \citet{1805.02016}.

As a final remark, note that, in some sense, the `box method' is conservative as it accounts for all possible correlations and at least some non-Gaussian features of the $(M,\astar)$ distribution.
This naturally leads to weaker exclusion limits, i.e.\ smaller exclusion regions for \ma, as can be seen in \cref{fig:comparison}.
However, the $(M,\astar)$ distribution contains additional information, and discarding it seems unnecessary.
Moreover, the `$2\sigma$ interval' relies on a (possibly ill-defined) extrapolation from the `$1\sigma$ error' of $\hat{M}$ and \astarhat.\footnote{Note that \citet{1411.2263} directly use the 95\% intervals for \astar instead of an extrapolation.}
Due to this, it is not straightforward to apply the `box method' to multiple BHs or to compare to limits obtained at a different exclusion level.

\subsection{Limitations and possible extensions}\label{sec:limitations}

As explored before, more powerful limits on ULB masses can be derived using higher SR levels.
While we can compute the free SR rates for $n > 2$, the most reliable limits on the self-coupling are currently only possible for $n = 2$.
One important extension of existing work is to compute the equilibrium rates when higher levels are relevant.
Furthermore, one may include more interactions, such as stimulated decay in the case of the axion-photon coupling $g_\gamma \propto \ifax$.

From the data modelling side, our method can also easily include additional parameters to propagate their uncertainties into the ULB exclusion regions.
One such parameter is the BH timescale \tauBH, as discussed in \cref{sec:timescales}, which could be included with a physically-informed prior distribution or constrained by related data sets.
We also did not consider possible interactions between the boson cloud and the SMBH environment, which could affect the cloud growth.
For instance, \updated{\citet[][see also \citealt{1411.2263}]{du_daniel_2022}} computed the boson cloud's perturbative stability in a thin accretion disc or in the presence of a stellar halo and found that their stability estimator worsens as $\alpha^q$ with $q \gtrsim 5$.  

Additional nuisance parameters can also be included when inferring the mass and spin of a given BH to better reflect their uncertainties.
In \cref{sec:bh_data}, we provided a summary of various aspects of the BH analysis to highlight the underlying assumptions and possible sources of uncertainties.
In the case of X-ray reflection modelling, several parameters, e.g.\ the inclination and iron abundance of the accretion disc, are well-known to be degenerate with the spin in state-of-the-art reflection models.
In addition, in the case of continuum-fitting models used in XRBs, spin inference is affected by intrinsic astrophysical uncertainties, such as the inclination angle, distance, or the BH's mass.
More information on these parameters could be provided by involving more constraints, such as data from current and upcoming X-ray polarimetry missions including the currently operating \textit{IXPE} observatory.

Moreover, in addition to including more parameters, we could also employ a hierarchical Bayesian analysis, considering both individual BH data as well as BH population parameters and relations.
This has already been explored in the context of BBH mergers (see \cref{sec:gw_bhs}), which we did not consider in this work.
The past work assumed hyperpriors for the initial masses and spins of the merging BH, which we could also employ for SMBHs and XRBs.
This would allow us to compute the final BH mass and spin from the full BH evolution including SR, which can be compared to the BH data \citep[potentially using methods from simulation-based inference; see e.g.][]{1911.01429}.
Finally, note that various nuisance parameters, such as the virial factor \colfac \citep[see e.g.][Fig.~2]{astro-ph/0407297} or a possible $M$-$L$ relationship between the BH mass and luminosity \citep[see e.g.][Figs~6--7]{1910.11875} might be exploited to refine the $(M,\astar)$ inference for BHs.
While the complexity of our analysis would increase when employing hierarchical model, these are natural extensions of a Bayesian framework, which -- apart from refining the accuracy of its inference -- could help to improve convergence and gain new insights into BH physics.

Compared to both alternative methods, our Bayesian statistical approach more easily generalises to a joint analysis of multiple BHs and the inclusion of other constraints: the Bayesian logic is naturally extendable while allowing us to include any correlations or additional uncertainties through nuisance parameters where necessary.
This is useful when considering BHSR constraints in global fits of axion models, such as the analysis by \citet{1810.07192} within the \code{GAMBIT} global-fitting framework.
Apart from previously used BH data \citep[see e.g.][]{2009.07206}, such a global fit could also consider the well-studied XRB GRS~1915+105, whose spin has been estimated from both the X-ray reflection and continuum-fitting methods \citep{mcclintock2006,soumya2020}.
Another promising prospect is the inclusion of future data on tidal disruption events near SMBHs from the Vera C. Rubin Observatory Legacy Survey of Space and
Time (LSST), which depend on the BH spin and can constrain ULBs in the range of $\SI{e-20}{\eV} < \ma < \SI{e-18}{\eV}$ \citep{du_daniel_2022}.

\section{Summary and conclusions}\label{sec:conclusions}

We have presented a universal approach for obtaining constraints on ultralight bosons (ULBs) from black hole (BH) data.
We exemplified our methodology with two well-studied BHs: \abhname, a stellar-mass BH, and \smbhname, a supermassive BH.
For each of these, we used posterior samples for the BH's mass and spin $(M,\astar)$, and computed the posterior probability on the ULB model parameters $(\ma,\ifax)$.
In particular, these are the first ULB constraints derived from \smbhname.

Our method introduces a rigorous statistical framework, extending previous approaches and preserving their advantages while, at the same time, giving more accurate or powerful constraints.
In particular, our method appears to be the only proposed methodology to self-consistently derive ULB constraints from superradiance (SR) in multiple BHs.
We only rely on samples from the $(M,\astar)$ posterior distribution alone and need not reproduce the entire BH data analysis tool chain.

Our approach also allowed us to clarify the statistical nature and reasons for disagreement between previous works (see \cref{sec:comparison}).
In particular, we improve on previous methods by capturing correlations and non-Gaussianities of the $(M,\astar)$ distribution, providing a more complete treatment of \emph{statistical} uncertainties from the BH analyses.
Still, we emphasise again that BH inference is affected by potentially large \emph{systematic} uncertainties related to the assumptions and limitations of accretion disc models, BH evolution, bosenova vs equilibrium scenarios, or computation of the SR rates.

Let us summarise the most important results from our work and the comparison to the literature.
In this initial study we observe that
\begin{itemize}
    \item the proposed Bayesian approach automatically includes all information from the data while being easily realisable,
    \item neglecting correlations between $M$ and \astar may be acceptable,
    \item the Gaussian approximation holds for $\ln(M/\Msol)$ instead of $M$ for SMBHs,
    \item the Gaussian approximation is poor when $|\astar| \sim 1$,
    \item the `box method' leads to overly conservative limits and does not leverage the full information contained in the BH data,
    \item inclusion of higher levels up to $n > 2$ can potentially constrain higher masses by up to an order of magnitude,
    \item compared to the bosenova, the equilibrium method for self-interactions reduces the maximum value of \ifax probed by any given BH by approximately one order of magnitude.
\end{itemize}

From the theoretical side, we are still limited by the availability of multi-level equilibrium BH evolution predictions, which would allow us to extend constraints for larger \ma values.
From the data side, a wider applicability of our method is limited by the availability of a (sampled) distribution of $(M,\astar)$, and we thus encourage all groups deriving these constraints to make their likelihoods or posteriors publicly available.
Each new data set will at worst linearly increase the computational time required for the sampling and more likely be sublinear due to faster convergence and due to abandoning parameter points that fall below some likelihood threshold after considering part of the data.
Overall, the computational effort should be manageable.

Our method has improved on the statistical interpretation of BHSR constraints compared to previous works \citep{1805.02016,2011.11646}, and we have identified causes of disagreement in the literature due to rates models and statistics.
Our conclusions have consequences for BHSR constraints on the string theory landscape.
\cite{2103.06812} computed constraints on the landscape using a version of the implementation of rates and statistics by \cite{2009.07206}.
We have verified the statistical approximation of \cite{1805.02016,2009.07206} as self-consistent and valid.
An error in the low mass end rates of \cite{1805.02016,2009.07206} is relatively small and should not affect conclusions about constraints on the landscape, due to the logarithmic nature of the \ma distribution.
Likewise, using the equilibrium rather than bosenova model for self-interactions will lead to only a slight shift in the maximum topological numbers probed by BHSR, due to the relatively small shift in $\ifax$ and slow trends in landscape models on the variation of this parameter.
On the other hand, inclusion of higher levels in \cite{2103.06812} significantly widens the \ma range that BHSR can probe.
The effect this will have on conclusions relating to the landscape is unclear.
\cite{2103.06812} used a large number of overlapping BHs in the stellar regime, so the effect from any single BH will be small, and the limits should only be affected by the inclusion of higher levels on the lowest mass BH in the sample.
Nonetheless, this underscores the importance of computing full BH Regge trajectories in future improvements of BHSR constraints, in order to leverage the constraining power of higher levels accurately, and the importance of extending our methods to large numbers of BHs if the posterior samples can be made available.

Possible extensions of our Bayesian method are discussed in detail in \cref{sec:limitations}, including additional parameters related to the BH timescale, Bayesian hierarchical modelling \updated{of BH populations, or a companion in binary systems such as \abhname.}
Assuming hyperprior distributions for the initial BH mass and spin would further allow us to compute the final BH mass and spin from the full SR evolution.
This treatment is more refined than simply comparing the associated SR and BH timescales.
Similarly, this approach can also incorporate the constraints from binary BH mergers, following \citet{1908.02312,2011.06010,2201.11338}.
We can also include the physics of direct gravitational wave emission from the boson cloud in our model, which has been considered by \citet{1604.03958,1706.06311,1804.03208,1909.08854,2111.15507,2404.16265}.

Moreover, constraints on $\ma$ and $\fax$ from BHSR are highly complementary to experimental and astrophysical probes of ULBs, with particular application to the QCD axion, fuzzy DM, and string theory.
Our systematic approach will allow this complementarity to be leveraged in global statistical analyses.

To further facilitate these, and possibly other, future extensions, we made our software code available on Github~\citep{github_bhsr}.

\section*{Acknowledgements}
We thank Masha Baryakhtar, Marios Galanis, Olivier Simon, and Sam Witte for helpful, substantial input on our draft and discussions related to their ongoing and past works; Lijun Gou and Xueshan Zhao for $(M,\astar)$ samples from \citet{1601.00615}; Jinyi Shangguan for \astar samples from the \citet{1705.02345,2009.08463}; Dominic J.\ Walton for \astar samples from \citet{walton2020}; and Jakob van den Eijnden, Rob Fender, and Matt Jarvis for helpful discussions on BH masses.

SH has received funding from the European Union's Horizon Europe research and innovation programme under the Marie {Sk{\l}odowska}-Curie grant agreement No~101065579.
DJEM is supported by an Ernest Rutherford Fellowship from the STFC, Grant No.~ST/T004037/1 and by a Leverhulme Trust Research Project (RPG-2022-145).
JSR acknowledges the support from the Science and Technology Facilities Council (STFC) under grant ST/V50659X/1 (project reference 2442592) and the Institute of Astronomy, as well as from a NASA Astrophysics Data Analysis Program (grant 80NSSC24K0617).
JHM acknowledges funding from a Royal Society University Research Fellowship.
We acknowledge the CINECA award under the ISCRA initiative, for the availability of high-performance computing resources and support.

We made use of the \code{BibCom}~\citep{bibcom} and \code{WebPlotDigitizer}~\citep{webplotdigitizer} tools, as well as the Python packages \code{numba}, \code{numpy}~\citep{10.1038/s41586-020-2649-2}, and \code{scipy}~\citep{10.1038/s41592-019-0686-2}.


\section*{Data Availability}
The Python software code and data underlying this article are available on Github at \url{https://github.com/sebhoof/bhsr}.


\bibliographystyle{mnras}
\bibliography{references}


\bsp 
\label{lastpage}
\end{document}